\documentclass[a4paper,fleqn]{cas-dc}

\usepackage[numbers]{natbib}
\usepackage{amsmath,amssymb}
\usepackage{algorithmic}
\usepackage{graphicx}
\usepackage{textcomp}
\usepackage{xcolor}
\usepackage{xspace}
\usepackage{tabularx}
\usepackage{hyperref}
\usepackage{balance}
\usepackage{xcolor}
\usepackage{booktabs}
\usepackage[inline]{enumitem}
\usepackage[justification=centering]{caption}
\usepackage{xcolor,colortbl}

\newcommand{\etal}{et al.\xspace}
\newcommand{\ie}{i.e.,\xspace}
\newcommand{\eg}{e.g.,\xspace}
\newcommand{\fig}[1]{Fig.~\ref{#1}}
\newcommand{\tab}[1]{Table~\ref{#1}}
\newcommand{\sect}[1]{Section~\ref{#1}}
\newcommand{\changed}[1]{{\color{red}#1}}
\renewcommand\changed[1]{#1}

\newcommand{\github}{GitHub\xspace}
\newcommand{\bitbucket}{BitBucket\xspace}
\newcommand{\gitlab}{GitLab\xspace}
\newcommand{\git}{git\xspace}
\newcommand{\python}{\textsf{Python}\xspace}

\newcommand{\bodegha}{\textsf{BoDeGHa}\xspace}

\def\hyphenateAndTtWholeString #1{\xHyphenate#1$\wholeString\unskip}

\def\xHyphenate#1#2\wholeString {\if#1$%
    \else\transform{#1}%
    \takeTheRest#2\ofTheString\fi}

\def\takeTheRest#1\ofTheString\fi
{\fi \xHyphenate#1\wholeString}

\def\transform#1{\url{#1}\hskip 0pt plus 1pt}

\def\urlx #1{\href{#1}{\hyphenateAndTtWholeString{#1}}}

\usepackage[]{todonotes}

\definecolor{myblue}{HTML}{3276AF}
\definecolor{myorange}{HTML}{F38636}

\def\tsc#1{\csdef{#1}{\textsc{\lowercase{#1}}\xspace}}
\tsc{WGM}
\tsc{QE}
\tsc{EP}
\tsc{PMS}
\tsc{BEC}
\tsc{DE}

\begin{document}\sloppy
\let\WriteBookmarks\relax
\def\floatpagepagefraction{1}
\def\textpagefraction{.001}
\shorttitle{Detecting bots in issue and PR comments}
\shortauthors{Mehdi Golzadeh et~al.}

\title [mode = title]{A ground-truth dataset and classification model for detecting bots in GitHub issue and PR comments}
\tnotemark[1]

\tnotetext[1]{This work is supported by the Fonds de la Recherche Scientifique -- FNRS [ Grants number T.0017.18, O.0157.18F-RG43 and J.0151.20 ]}

\author[1]{Mehdi Golzadeh}[auid=000,bioid=1,
                        orcid=0000-0003-1041-439X]
\cormark[1]
\fnmark[1]

\address[1]{Software Engineering Lab, University of Mons, Avenue Maistriau 15, 7000 Mons, Belgium}
\author[1]{Alexandre Decan}[orcid=0000-0002-5824-5823]
\fnmark[2]

\author[1]{Damien Legay}[orcid=0000-0001-6811-6585]
\fnmark[3]

\author[1]{Tom Mens}[orcid=0000-0003-3636-5020]
\fnmark[4]

\cortext[cor1]{Corresponding author}
\fntext[fn1]{mehdi.golzadeh@umons.ac.be}
\fntext[fn2]{alexandre.decan@umons.ac.be}
\fntext[fn3]{damien.legay@umons.ac.be}
\fntext[fn4]{tom.mens@umons.ac.be}

\begin{abstract}
Bots are frequently used in Github repositories to automate repetitive activities that are part of the distributed software development process. They communicate with human actors through comments.
While detecting their presence is important for many reasons, no large and representative ground-truth dataset is available, nor are classification models to detect and validate bots on the basis of such a dataset.
This paper proposes a ground-truth dataset, based on a manual analysis with high interrater agreement, of pull request and issue comments in 5,000 distinct Github accounts of which 527 have been identified as bots.
Using this dataset we propose an automated classification model to detect bots, taking as main features the number of empty and non-empty comments of each account, the number of comment patterns, and the inequality between comments within comment patterns.
We obtained a very high \changed{weighted average precision, recall and F1-score of 0.98} on a test set containing 40\% of the data. %
We integrated the classification model into an open source command-line tool to allow practitioners to detect which accounts in a given Github repository actually correspond to bots.
\end{abstract}

\begin{keywords}
distributed software development \sep bot identification \sep GitHub repositories \sep text similarity \sep classification model
\end{keywords}

\maketitle

\section{Introduction}
\label{sec:intro}

The collaborative nature of software development has inherently made it a social phenomenon, which has led to the advent of {\em social coding} platforms such as GitHub, BitBucket, and GitLab~\cite{DabbishSTH12}.
These online platforms have taken the collaborative nature of open source software development to a new level, by integrating mechanisms such as issue reporting, pull requests (PR), commenting and reviewing support into distributed version control tools~\cite{Gousios2014,Tsay2014,Tsay2014LTE}.
The pull-based development process is the primary means for integrating code from thousands of developers in distributed development platforms such as \github~\cite{Gousios2014}.
This model has had a significant impact on the development of open-source software, but at the same time has significantly increased the workload of repository maintainers to communicate with other contributors, review source code, deal with contributor license agreement issues, explain project guidelines, run tests and build code, and merge pull requests~\cite{Gousios2016}.

To reduce this workload, developers have been adopting automated tools to perform repetitive tasks in the development process~\cite{Wessel2018}, such as updating dependencies \cite{Mirhosseini2017} (e.g. \emph{dependabot}) and fixing vulnerabilities (e.g. \emph{snykbot}), improving code reviews (e.g. \emph{Review bot})~\cite{Balachandran2013} and documenting code refactorings \cite{Rebai2019}.
Such tools are commonly known as {\em DevBots}~\cite{Erlenhov2020ESECFSE}, or {\em bots} for short. They are generally seen as a promising approach to deal with the ever-increasing complexity of contemporary distributed software development.

While the use of bots in open source software repositories can alleviate maintainer workload, their presence poses challenges for empirical software engineering researchers that aim to study socio-technical aspects of software development. For example, in a previous study we analysed the impact of discussions on pull request (PR) decisions in \github repositories~\cite{Golzadeh2019} by studying these discussions in 188K PRs of \github repositories. We ignored the presence of bots in that study, deferring it to future work.
Repeating the same analysis taking into account the bots allowed us to discover that 20\% of those comments belong to bots, and that bots were involved in 31\% of all PRs. Bots were responsible of accepting or rejecting 25\% of all PRs. Moreover, we found that the proportion of successfully integrated PRs was twice as high for PRs involving bots.

Other empirical socio-technical analyses based on historical software repository data
are likely to have been biased as well by not considering the presence of bots.
Some empirical studies explicitly acknowledge the presence of bots, and attempt to remove them during data preprocessing (e.g. filtering out bots) or postprocessing (e.g. removing outliers)~\cite{Dey2019PROMISE19, Liu2019}.
It is therefore important to consider the presence of bots in such studies, and to treat them differently than humans.

A prerequisite for considering bots is the ability to identify their presence in software development activities. This is not a simple task because, depending on the considered data source, bots often do not have a distinct representation in social coding platforms, and may look, act like or even impersonate humans.
Our review of the research literature (see \sect{sec:related}) revealed a few attempts to manually identify and classify bots. We only came across one study attempting to automate the bot identification process based on commit activity in \github repositories~\cite{Dey2020MSR}.
The current paper has a similar focus, but based on a different data source, namely PR and issue comments in \github repositories.

As a first major contribution, we propose a large and reliable ground-truth dataset, consisting of 5,000 distinct \github accounts of which 527 were manually identified as bots based on their PR and issue commenting contents. As a second major contribution, we use this ground-truth to create and evaluate a classification model that relies on comment-related features to accurately classify accounts as either bot or human. As a third contribution, we propose an open-source tool based on the classification model to allow \github contributors to detect which accounts in their repositories actually correspond to bots.

The remainder of this paper is structured as follows:
\sect{sec:related} presents the related work.
\sect{sec:extraction} explains the steps to create the ground-truth dataset.
\sect{sec:featureselection} details which features we selected for the classification model.
\sect{sec:model} explains the workflow to select an appropriate classification model and evaluates the selected classification model.
\sect{sec:tool} presents \changed{an open source tool} implementing this model.
\sect{sec:discussion} discusses the results.
\sect{sec:threats} presents the main threats to validity of the research.
\sect{sec:futurework} outlines future research avenues and
\sect{sec:conclusion} concludes.\section{Related Work}
\label{sec:related}

The earliest idea of computer software imitating humans dates back to the ideas by Alan Turing in 1950~\cite{Turing1950}.
In recent years, the development of AI and machine learning has led to a proliferation of automated tools that substitute humans to perform particular repetitive tasks~\cite{Dale2016}. For example, chatbots imitate natural language to communicate with humans through a conversational interface~\cite{Lebeuf2018}, and automated social actors (ASA) automatically create content on social networks~\cite{Abokhodair2015}. Bots are also widely used in other contexts such as
education~\cite{Kerry2009,Benotti2014,Luke2017},
e-commerce~\cite{BENMIMOUN2017,Thomas2016}, customer services~\cite{GnewuchMM2017,Jain2018}, peer production communities such as Wikipedia~\cite{Cosley2007,Geiger2013},
\changed{and social networks (such as Twitter). Social network bots are generally aimed at spamming and sending fake news and hence they seek to hide their true nature. Because of this, numerous studies are focused on identifying them~\cite{Minnich2017,Efthimion2018,RODRIGUEZRUIZ2020,Amleshwaram2013}. These studies have proposed machine learning models which aim to identify bots based on features such as profile specification (\eg age, location and biography), tweet content (\eg hashtags, URLs and similarity of sentiments), tweet time (\eg burstness and average tweets per day), and user network profile (\eg interaction between users). Such features either do not have any equivalent in social coding platforms (\eg hashtags) or require a lot of effort to collect (\eg user network profiles). Moreover, bots in social networks appear to be quite different from bots in social coding platforms (\ie DevBots), whose purpose is to help developer teams carry out automated activities in the software development process. We did not find any evidence of intentionally malicious use of DevBots.}

In the context of software development, bots are automated software agents that perform repetitive well-defined tasks that support and integrate with the activities of human developers~\cite{Wessel2018, Farooq2016}.
They are capable of communication and decision making~\cite{Storey2016} and carry out tasks that involve interactions with humans~\cite{Lebeuf2017a}.
They support both technical and social activities~\cite{Lin2016} to coordinate collaborative software development~\cite{Perez-Soler2017}, such as
improving feedback on code contributions~\cite{Hu2019}, repairing continuous integration build failures~\cite{Urli2018}, and deployment and evaluation of software engineering analysis techniques~\cite{Beschastnikh2017}.

Recent research has focused on the practical value of bot adoption in software engineering, such as how bots increase software development productivity~\cite{Storey2016}, how bots enable faster software dependency updates~\cite{Mirhosseini2017} and how bots can help reduce the friction points software developers face when working collaboratively~\cite{Lebeuf2017}.
Other studies have introduced new bots and analysed their effect on software repository activities such as test bots~\cite{Erlenhov2020}, bots to improve newcomers' experience and help them to better engage in the project~\cite{Dominic2020},
bots for answering developer questions using historical Q\&A data~\cite{Romero2020}, bots for assisting in the development of microservice architecture and the use of NLP~\cite{Lin2020}.

A prerequisite for studying the impact of bots on software production processes is the ability to identify such bots in the first place. We found very few studies trying to identify and categorise bots. Wessel \etal~\cite{Wessel2018} conducted a study about prevalence and effect of bots in \github repositories. They manually analysed 351 repositories and found that 26\% of them use bots. By manual inspection of \github accounts they identified 48 different bots in 93 projects. They found statistical differences regarding the number of commits, number of changed files, and closing time of PRs between projects before and after bot adoption. They reported both positive and negative challenges of bot adoption from integrators and contributors' viewpoints.
In another study, they discuss six useful bots in \github's PR process~\cite{Wessel2020botse}. They analysed the negative aspects of bots in code contributions and introduce a meta-bot that acts as a middleman to mitigate this effect.

Erlenhov \etal~\cite{Erlenhov2019} presented a taxonomy that classifies 11 existing development-related bots in \github and Slack. Lebeuf~\cite{Lebeuf2018thesis} provided a multi-faceted classification of bots (including many well-known examples of bots), combining their properties and behaviour.
None of these studies proposes an automated approach to identify bots.

Dey \etal~\cite{Dey2020MSR} did propose an automatic method to identify bot accounts in \git projects.
Each identity in their dataset consists of an author name and email address.
They studied three different approaches to find bots, based on
\changed{
\begin{enumerate*}[label=(\roman*)]
    \item the presence of the string ``bot'' at the end of the author name,
    \item commit messages, and
    \item features related to files changed in commits and projects the commits are associated with.
\end{enumerate*}
They combined these three different approaches into a single ensemble model that was validated on a dataset of 67 bots of which 58 cases (85\%) were effectively captured by the model.}

Their study is fundamentally different from ours, since their dataset is based on commit data in \github repositories, whereas we will focus exclusively on \github issue and PR comments. They also identified authors based on the author name and email address, whereas we rely on the \github account name exclusively.
Both datasets are quite complementary, as we found many examples of bots that are only involved in commit activity and others that are only involved in issue and PR activities. Moreover, the nature and contents of commit comments is quite different from issue and PR comments, requiring other features to establish an accurate classification model.
\section{Ground truth dataset}
\label{sec:extraction}

In order to be able to evaluate an automated algorithm to detect bots based on their commenting activity in \github issues and pull requests, a ground truth dataset is required.
Such a ground truth dataset indicates, given a contributor commenting in an issue or a pull request, whether this contributor is a human or a bot.
To be effective and representative, the ground truth dataset should be large enough, \ie it should cover a considerable number of \github repositories, contributors, issues and pull requests.

Since we did not encounter any such representative ground truth dataset in the research literature, we set out to create it ourselves.
To do so, we downloaded and manually examined comments from thousands of issues and pull requests, labelling each contributor either as a bot or a human commenter.
Despite the considerable effort needed to create such a dataset, it was a worthwhile endeavour, since it will be a valuable resource for other researchers as well.

This section explains how we proceeded to create and validate our ground truth dataset, from the raw data we downloaded to the process of rating and labelling each contributor.

\subsection{Terminology}
\label{sec:terminology}

In the context of this paper, we will consistently use the following terminology.
We use the term \textbf{\emph{bot}} to refer to a \github bot, defined by Wessel~\cite{Wessel2018} as ``\emph{a task-oriented bot, responsible for automating well-defined tasks on \github repositories. A \github bot behaves like a human user, serving as an interface between users and services.}''

Since our study focuses on distributed software development on \github, we use the term \textbf{\emph{repository}} to refer to a \github repository. Contributors to a repository can be identified by their unique (\github) \textbf{\emph{account}}.
Contributions to a repository can take different forms, such as code \textbf{\emph{commits}}, \textbf{\emph{issues}} and \textbf{\emph{pull requests (PR)}}.
The focus of this paper will be on issues and PRs.

Contributors can add (uniquely identifiable) \textbf{\emph{comments}} to PRs and issues in a repository. We use the term \textbf{\emph{commenter}} to refer to the \github account having provided this comment.  We also use the term comment to refer to its actual textual content.
Since a commenter can be either a bot or a human contributor, we will refer to them as \textbf{\emph{bot commenter}} and \textbf{\emph{human commenter}}, which we will abbreviate to \textbf{\emph{bot}} and \textbf{\emph{human}}, respectively.

\subsection{Data extraction}
\label{sec:data-extraction}

Our goal is to identify bot and human commenters based on the comments they made in issues and pull requests of collaborative software development repositories on \github.
\github is one of the leading online collaborative development platform. As of \changed{November} 2020, \github reported having over \changed{48} million users and more than \changed{195} million repositories (including at least \changed{37} million public repositories).

Following the guidelines provided by Kalliamvakou \etal~\cite{Kalliamvakou2014}, we want to avoid repositories that have been created merely for experimental or personal reasons, or that only show sporadic traces of issue and PR comments.
Moreover, since our focus is on software development repositories, we want to exclude repositories that are not related to software development.
To comply with these constraints, we relied on \textsf{libraries.io}~\cite{Katz2020}, a monitoring service indexing information for several million packages being distributed through 37 software package registries, such as \textsf{npm}, \textsf{PyPI}, etc.

We downloaded the data dump of January 2020\footnote{Version 1.6 on \url{http://doi.org/10.5281/zenodo.3626071}} containing, among others, links to the \github repositories related to these distributed software packages. Since it contains more than 3.3 million \github repositories, we randomly selected around 136K of them as the starting point of our dataset creation process.
For each of these repositories, we extracted \changed{on 16 February 2020} %
 the last 100 comments of the last 100 issues and pull requests using \github's GraphQL API.
This resulted in over 10 million comments \changed{covering a period of more than 10 years (ranging from 17 December 2009 to 15 February 2020). These comments were made by} more than 837K distinct contributors, corresponding to more than 3.5 million issues and pull requests.
The extracted comments also include the textual description of each considered PR. While the \github API does not consider PR descriptions as comments, we do, since the \github web interface does not visually distinguish them from other comments.

Since our goal is to distinguish between bots and human contributors based on their comments, we require a sufficiently large number of comments for each commenter. Hence, we decided to exclude commenters who made fewer than 10 comments based on a threshold we identified in a previous study~\cite{Golzadeh2020}.
At this stage of the process, the dataset contains 6,307,489 comments belonging to 79,342 contributors, spanning 42,492 repositories.

Since this is too much data to process manually, we extracted a subset covering  \changed{5,082} commenters.
\changed{This subset was composed of 4,644 randomly selected commenters to which we manually added 438 extra commenters that are more likely to correspond to bots based on previous studies~\cite{Golzadeh2020,Wessel2018} (52 cases), or because they contained a specific substring in their \github account name (386 cases).} The substrings we considered were ``bot'', ``ci'', ``cla'', ``auto'', ``logic'', ``code'', ``io'' and ``assist''.
\changed{By doing so, we increased the likelihood of having a sufficient number of bots in the dataset.}

The resulting subset contains 5,082 commenters and covers 3,975 repositories, 186,991 issues and pull requests, and contains 301,557 comments.
\tab{tab:dataextraction} summarizes the main characteristics of the considered datasets.

\begin{table}[!tb]
    \centering
    \caption{Summary of the dataset characteristics.}
    \label{tab:dataextraction}
    \begin{tabular}{l|r}
        \toprule
        \bf raw dataset & \bf number  \\
        \midrule
        \github repositories  & 136,529   \\
        \quad $\hookrightarrow$ from \# distinct owners & 84,983 \\
        issues   & 1,588,363   \\
        pull requests (PR)      & 1,951,705  \\
       issue and PR comments & 10,874,611  \\
       \quad $\hookrightarrow$ from \# distinct commenters & 873,489 \\
        \bf selected subset\rule{0pt}{3ex} &  \\
        \midrule
        \github repositories  & 3,975    \\
       \quad $\hookrightarrow$ from \# distinct owners & 3,425\\
        issues & 50,241  \\
        pull requests (PR) & 136,750   \\
        issue and PR comments & 301,557    \\
       \quad $\hookrightarrow$ from \# distinct commenters & 5,082 \\
        \bottomrule
    \end{tabular}
\end{table}

\subsection{Data labelling and rating process}
\label{sec:rating}

The next step to create a ground truth dataset is to manually identify bots and humans.
To ease this process, we developed a web application through which \changed{the list of comments of each commenter was presented to at least two raters (among the four authors of this paper)}.
Comments were displayed by batches of 20, \changed{starting with the most recent comments first,} and the rater had an option to display more comments if needed.
\changed{The account name of the commenter was not revealed to avoid bias, as the goal was to classify commenters based on their comments only.}
The rater could select whether the commenter is considered as a ``Bot'' or a ``Human''. In case a rater was uncertain whether the commenter was a bot of a human being, a third option could be selected: ``I don't know''.
Furthermore, the rater was asked to select a difficulty level among ``Very easy'', ``Easy'', ``Difficult'' and ``Very difficult'' for his decision.

\begin{figure*}[!tb]
    \centering
    \includegraphics[scale=0.33]{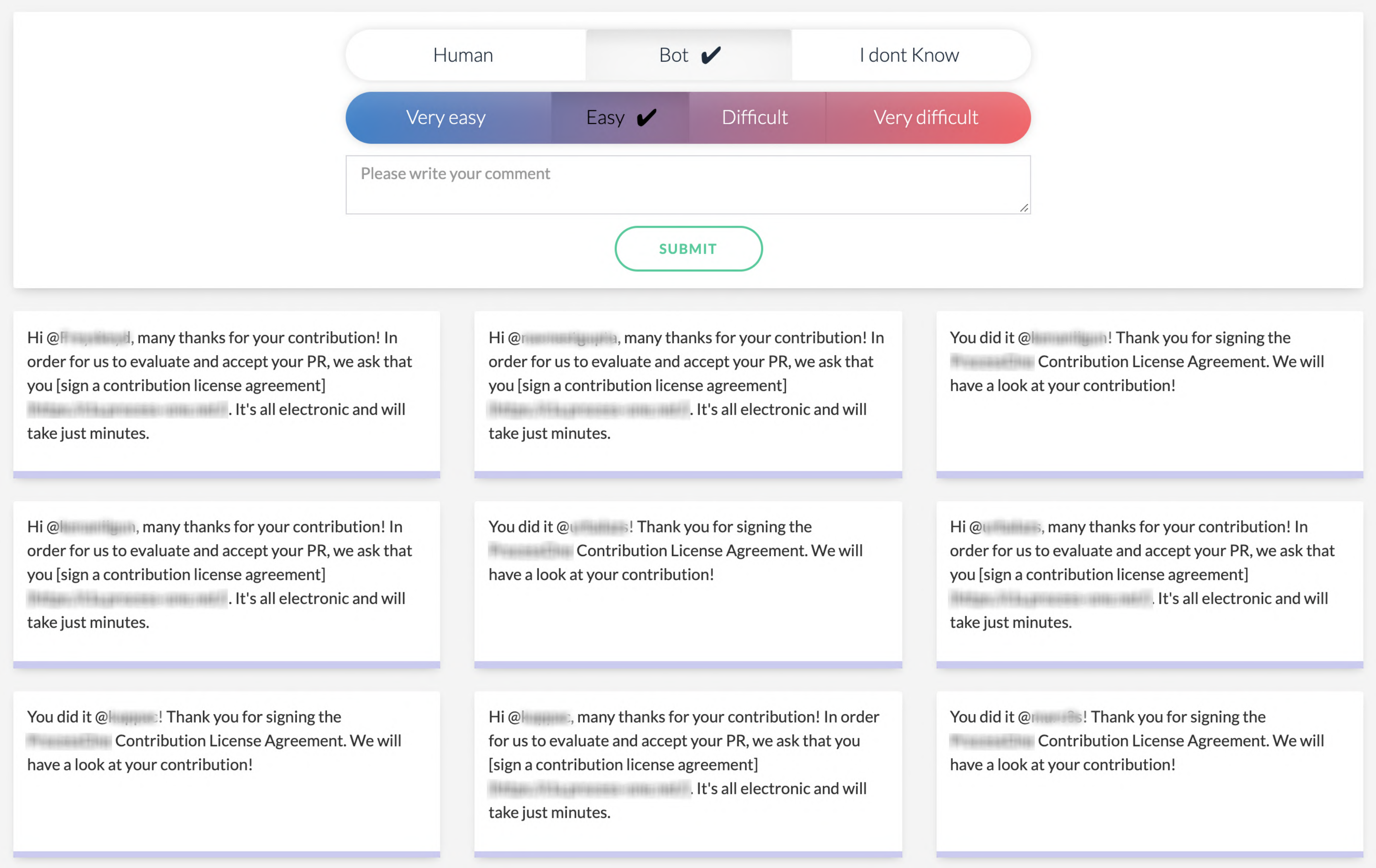}
    \caption{Anonymised screenshot of the rating application in action.}
    \label{fig:ratingapp}
\end{figure*}

\fig{fig:ratingapp} shows a screenshot of the rating application in action.
For the specific example being shown, raters could easily decide that the commenter is a bot based on the content and repetitiveness of all visible comments.

\begin{figure}[!tb]
    \centering
    \includegraphics[scale=0.5]{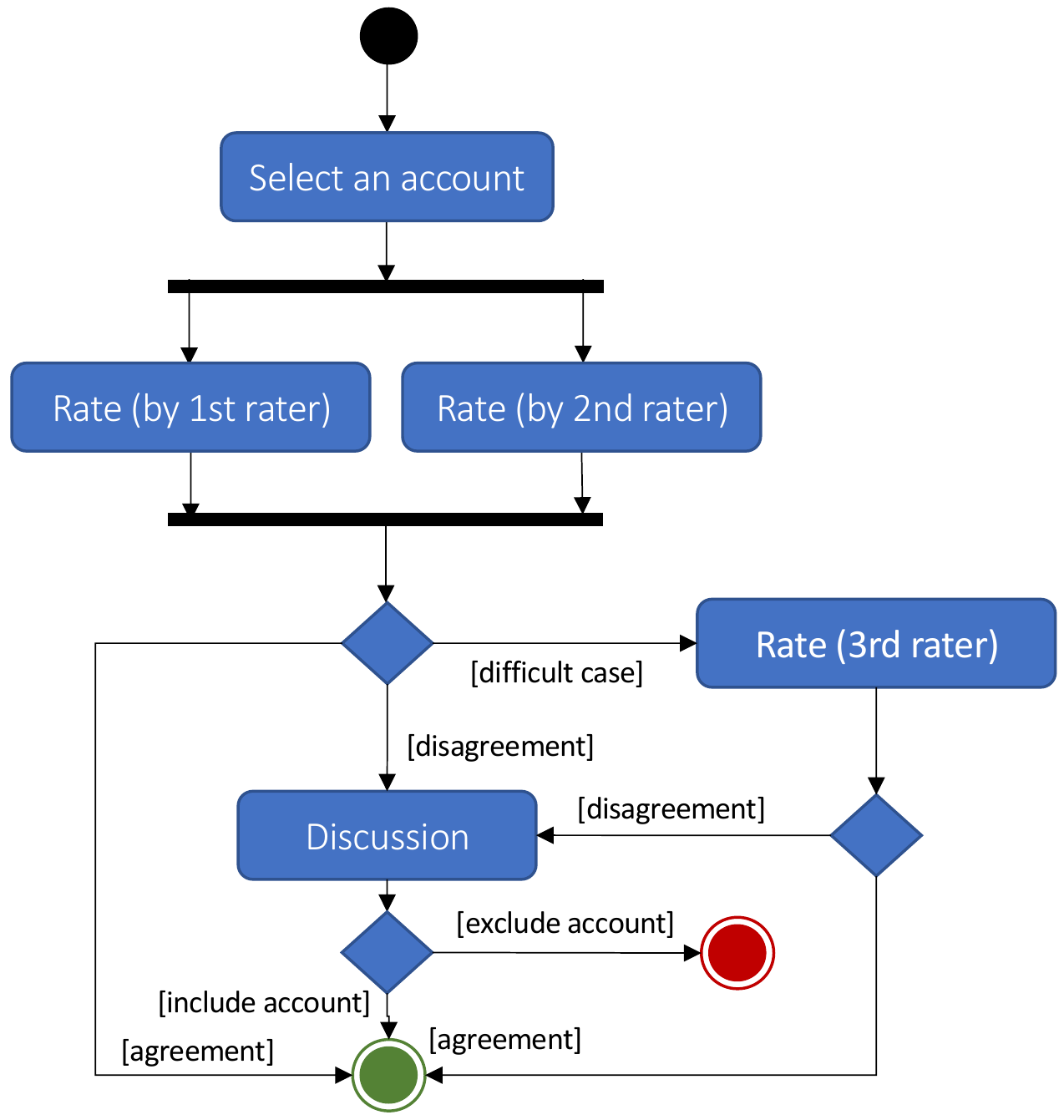}
    \caption{Workflow of the rating process.}
    \label{fig:ratingflow}
\end{figure}

In total, 5,082 commenters were rated, ending up with exactly 5,000 commenters after having filtered out 82 commenters during the following process.
The rating process was performed in two steps to come with an optimal inter-rater agreement, relying on Landis agreement levels~\cite{Landis77}.
The rating process is summarized in \fig{fig:ratingflow}.
Each commenter was initially rated by two distinct raters.
All cases that were agreed either as bot or human were included in the ground-truth dataset.
In order to assess the reliability of the ground-truth dataset, we computed the inter-rater reliability (IRR)~\cite{campbell2013-IRR} between each pair of ratings based on Cohen's kappa $\kappa$~\cite{mchugh2012interrater}. The results are presented in \tab{tab:ratingsummary}.

\begin{table}[]
    \centering
    \caption{Summary of two-step rating process.}
    \label{tab:ratingsummary}
    \begin{tabular}{lrr}
    \toprule
     & \bf first & \bf second  \\
     & \bf  \changed{step} & \bf \changed{step} \\

    \midrule
    commenters agreed as \emph{bot} &  472 &  527  \\
    \quad $\hookrightarrow$ from \# repositories & 457 & 505 \\
    commenters agreed as \emph{human} &  4,364 & 4,473 \\
    \quad $\hookrightarrow$ from \# repositories & 3,425 & 3,515 \\
    proportion of bots & 9.8\% & 10.5\%\\
    \midrule
    commenters agreed as ``I don't know'' & 69 & -- \\
    commenters without agreement &  177 & 4 \\
    commenters agreed as ``mixed'' & -- & 78 \\
    \midrule
    $\kappa$ agreement score & 0.84 & 0.96 \\
    \bottomrule
    \end{tabular}
\end{table}

The first step of the rating process ended up with 472 bots and 4,364 humans, with a ``\emph{substantial}'' agreement ($\kappa=0.84$) between raters. At the end of this step, there were 246 cases because they were either not agreed (177 cases) or agreed as ``I don't know'' (69 cases). Additionally, 91 cases evaluated as ``difficult'' or ``very difficult'', leading to a total of 268 cases for the second step.

In a second step, we involved a third rater for the cases that were identified as ``difficult'' or ``very difficult'' during the first step. We then discussed all together all cases for which an agreement could not be achieved,  or the cases where the third rater disagreed with one of the two former ones.
During these discussions, we sometimes relied on additional information (\eg we looked at the \github account of the commenter, at time intervals between comments, the overall activity of the account, etc.) to come to a decision.

The large majority of cases we discussed were resolved on the basis of an unanimous decision between raters, leading to an ``\emph{almost perfect}'' inter-rater reliability ($\kappa=0.96$).
At the end of the second step, only 82 cases were left out of the ground-truth dataset, either because no agreement could be reached (4 cases), or because we agreed on the ``mixed'' nature of these commenters. These ``mixed'' commenters correspond to human commenters that relied on automatic tools to generate comments, therefore ``mixing'' the behaviour of a human and a bot at the same time.

For example, some of these accounts rely on an automated tool to facilitate code review by sending PRs to \emph{Reviewer}, a code review tool for GitHub. Other examples include the use of tools such as \emph{StyleCI} to improve code style, or \emph{semantic-release} to automatically determine the next version number of a release, generate release notes and publish a package.
We will discuss these ``mixed'' commenters in more details in \sect{sec:discussion}.

This left us with 5,000 commenters, of which 527 (\ie 10.5\%) are bots.
\tab{tab:groundtruth} summarizes the characteristics of final ground-truth dataset.
Since we believe such a ground-truth dataset is valuable for the research community (\eg to have a list of known bots, to study their characteristics or to train other models), we share it publicly on \url{http://doi.org/10.5281/zenodo.4000388}.
This dataset contains the name of the repository, the name of the commenter and whether it is a bot or a human.
Due to GDPR regulations and in order to protect \github users' privacy, we do not provide additional information (\eg their comments).

\begin{table*}[]
    \centering
    \caption{Summary characteristics of final ground truth dataset.}
    \label{tab:groundtruth}
    \begin{tabular}{lrrr}
    \toprule
    \bf number of... & \bf bot & \bf human & \bf total \\
    \midrule
    commenters & 527 & 4,473 & 5,000 \\
    repositories with at least 1 commenter & 505 & 3,515 & 3,909  \\
    comments  &  28,287 & 268,504 & 296,791 \\
    issues with at least 1 commenter & 2,749 & 46,959 & 49,623 \\
    PRs with at least 1 commenter & 16,937 & 118,896 & 134,208 \\
    \bottomrule
    \end{tabular}
\end{table*}

\section{Feature selection}
\label{sec:featureselection}

In this section, we explain the features that will be used by the classification model to distinguish bots from human commenters. These features include \changed{the} number of comment patterns, the number of (empty) comments, and the number of comments within each pattern.
The following subsections explain these features and the rationale behind their selection.

\subsection{Text distance between comments}
\label{sec:distance-metrics}

Based on the assumption that bots perform more repetitive and automated tasks, we hypothesise that bot commenters exhibit more repetitive comments than human commenters.
Consequently, we expect comments belonging to a bot to exhibit more similarity than comments belonging to a human commenter.
In order to measure the similarity between comments of each commenter, both in terms of content and structure, we rely on text distance metrics that are commonly used for this purpose in natural language processing.
The two metrics we consider are the Jaccard~\cite{Jaccard1912} and Levenshtein~\cite{Levenshtein1966} distances. The first one aims to quantify the similarity of two texts based on its content, and the second one captures the structural difference by counting single character edits.

More precisely, the Jaccard distance $J(C_1, C_2)$ measures the distance between two texts $C_1$ and $C_2$ by comparing the number of distinct common words in $C_1$ and $C_2$ with the total number of distinct words in $C_1$ and $C_2$.
If $words(C)$ denotes the set of words in $C$, then $J(C_1, C_2)$ is computed as:
$$\mathcal J(C_1, C_2) = 1 - \frac{\mid words(C_1)\cap words(C_2) \mid}{\mid words(C_1)\cup words(C_2)\mid}$$

The second distance we consider is the Levenshtein edit distance $lev(C_1,C_2)$ that measures the difference between two character sequences $C_1$ and $C_2$ by counting the minimum number of single-character edits (insertion, deletion, or substitution) required to convert $C_1$ into $C_2$.
We rely on its normalized version, computed as:
$$\mathcal L(C_1,C_2) = \frac{lev(C_1,C_2)}{max(|C_1|,|C_2|)}$$

To support our assumption that comments made by a bot have higher similarity than comments made by a human, we computed for each commenter in the ground truth dataset the Jaccard and Levenshtein distances between all pairs of comments belonging to that commenter.
In order to compute the Jaccard distance, we first needed to split comments into words, a process also known as \emph{tokenization}. To do so, we relied on \textsf{spaCy}, an ``industrial-strength natural language processing library''\footnote{\url{https://spacy.io}} that notably offers a fast but robust tokenization algorithm, among others.

\begin{figure}[h]
	\centering
	\includegraphics[width=\columnwidth]{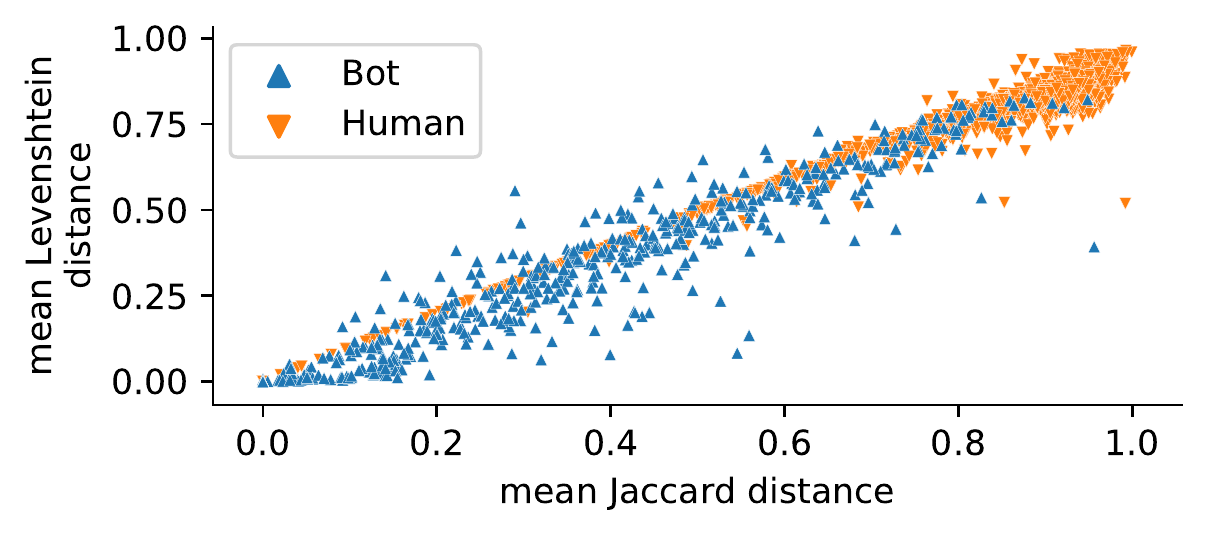}
	\caption{Mean Levenshtein and Jaccard distances between pairs of comments, per commenter.}
	\label{LevJaccard}
\end{figure}

\fig{LevJaccard} shows the mean Levenshtein and Jaccard distances for each commenter, distinguishing between bots (blue triangles) and humans (orange triangles).
We observe that many humans are grouped in the top right part of the figure, \ie they have high mean values for both distances. On the other hand, most bots have lower values for their mean distances. For instance, 91.6\% of all bots have mean Jaccard and Levenshtein distances below 0.75. For comparison, only 7.2\% of all human commenters exhibit mean Jaccard and Levenshtein distances below 0.75. %

Despite this, there is still a lot of overlap between bots and humans in \fig{LevJaccard}, indicating that the mean distances are not enough to properly distinguish between bots from humans.
By manually inspecting the comments belonging to bots having high mean distances, we found that their comments usually form sets of similar comments. Even if the distance between comments in a set (\ie intra-set distance) is low, the distance between comments belonging to different sets (\ie inter-set distance) is high. As a consequence, the overall mean distances between all comments tends to remain high, rivalling the distances observed for most human commenters.

We found many of these cases. One example is the bot that was identified in \fig{fig:ratingapp}. We observe that it has two different sets of similar comments.
The first set consists of comments of the form ``\emph{You did it @\ldots! Thank you for signing the \ldots Contribution License Agreement.
We will have a look at your contribution!}''.
The second set consists of comments of the form ``\emph{Hi @\ldots, many thanks for your contribution! In order for us to evaluate and accept your PR, we ask that you [sign a contribution license agreement] \ldots It's all electronic and will take just minutes.}''.
The mean distance between pairs of all 20 comments belonging to the first set (\ie intra-set distance) is very low (0.06 and 0.08 for Levenshtein and Jaccard distance respectively) and even lower (0.04 and 0.05 respectively) for the second set of 27 comments. However, the intra-set distance (\ie the distance obtained by comparing comments from the first pattern with comments for the second pattern) is much much higher (0.70 and 0.81 for Levenshtein and Jaccard distance respectively).
Consequently, the overall mean distances between all pairs of comments are 0.37 for Levenshtein and 0.43 for Jaccard distance. These distances are usually observed for human commenters, not for bots.

\subsection{Repetitive comment patterns}
\label{sec:clustering}

Since high mean distances between comments of a commenter could correspond to either a human or, in many cases, to a bot having sets of similar comments, we cannot exclusively rely on these mean distances to distinguish between bots and humans.
However, we observed that bots tend to have sets of many similar comments (\ie they follow comment patterns), while we found that most comments from humans are unique and only a few of them seem to follow a pattern (\eg ``\emph{Thank you!}'', ``\emph{LGTM}''\footnote{Shorthand for ``Looks Good To Me'', a common way among \github users to agree with what is proposed in a pull request.} or ``\emph{+1}''\footnote{This is another common way of expressing agreement with what was proposed in the previous comment or in the issue or PR description.}).

Based on this observation, we expect bots to have a lower number of comment patterns than humans. In order to capture these comment patterns, we rely on a clustering algorithm.
Clustering aims to group items into sets (``clusters''), in such a way that items belonging to the same cluster are more similar than items belonging to different clusters.

We selected \textsf{DBSCAN} (Density Based Spatial Clustering of Applications with Noise)~\cite{Ester1996DBSCAN}, a well-known density-based clustering algorithm that notably has the ability (i) to generate clusters of unequal size (\ie we can have patterns with unequal numbers of comments), (ii) to generate a single cluster if needed (\eg a commenter whose comments are all the same), and (iii) to generate single item clusters (\eg a commenter whose comments are all very different). Additionally, \textsf{DBSCAN} permits not to specify the number of clusters in advance, fitting our use case wherein we do not know the number of patterns of each commenter in advance.

Since we aim to capture both the structural and content distance between comments, we rely on a combination of the Levenshtein and Jaccard distance, defined as follows: $$\mathcal D(c_1,c_2) = \frac{\mathcal L(c_1, c_2) + \mathcal J(c_1, c_2)}{2}$$

\begin{figure}[h]
	\centering
	\includegraphics[width=\columnwidth]{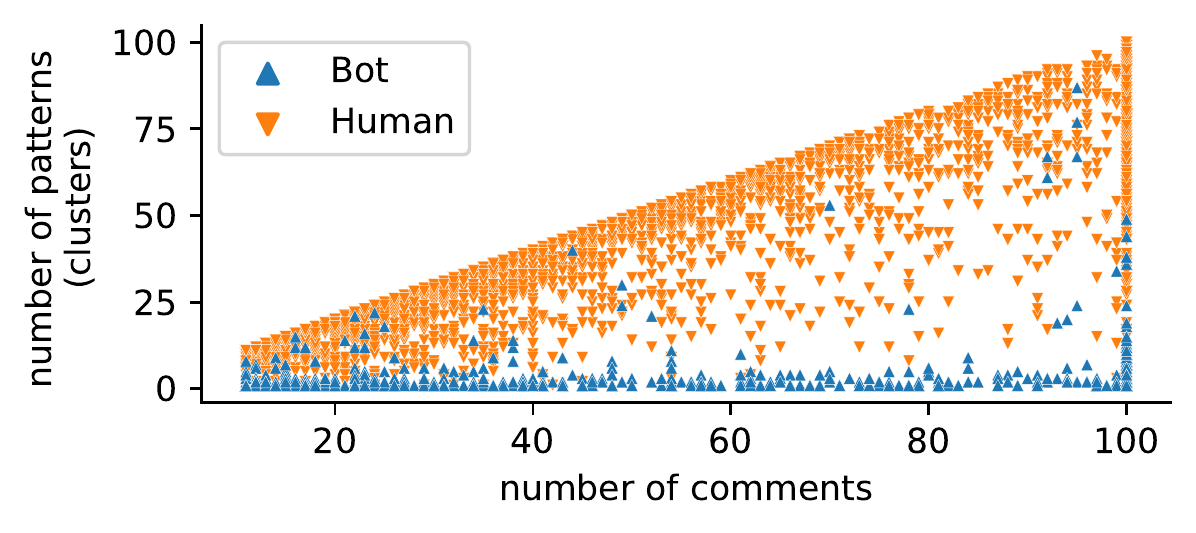}
	\caption{Number of comment patterns (clusters) and number of considered comments per commenter.}    \label{ClustersCMS}
\end{figure}

\changed{For each commenter, we computed $\mathcal D(c_i,c_j)$ for each pair $(c_i,c_j$) of comments. The resulting distance matrix, one per commenter, is then passed to \textsf{DBSCAN} to group the comments based on their similarity.}
\fig{ClustersCMS} reports on the number of patterns (\ie clusters), distinguishing between bot and human commenters.
Since the number of patterns could depend on the number of comments, we report on the number of patterns relative to the number of considered comments.

Compared to \fig{LevJaccard} we can observe a much clearer separation between bots and humans based on the number of comment patterns and the number of comments, although it is not perfect.
We observe that most humans are along the diagonal line which indicates that the number of patterns is close to the number of comments, and that almost all bots are along the horizontal axis. This means that the number of comment patterns for bots remains stable, and low, regardless of the number of comments they made.
This confirms our assumption that bots have a limited set of comment patterns, contrarily to humans that seems to make much more varied comments.

\subsection{Inequality between comments in patterns}
\label{sec:dispersion}

Although we expected human comments to be mostly non-repetitive (\ie each comment corresponds to a different pattern), we found instances in which a human commenter had a non-negligible number of repetitive comments (\eg ``\emph{Thank you!}'', ``\emph{LGTM}'' or ``\emph{+1}'') alongside other messages.
This leads to having human commenters whose number of comment patterns is much lower than the number of comments, which is exactly the assumption we had for bots due to their repetitive comments.
However, we found that those human commenters correspond to cases having at the same time a few patterns with many comments and many patterns with a few (mostly single) comments. On the other hand, bots exhibit single comment patterns less often.
For instance, among the 2,431 patterns corresponding to bots, 50\% are composed of a single comment, while this proportion is much higher (95.9\%) for the 230,711 patterns we have for humans.

This observation lead us to consider the inequality in the number of comments in each pattern as a supplementary feature to distinguish between bots and humans.
The Gini coefficient \cite{Dorfman1979Gini} provides a way to quantify the inequality (\ie the distribution) of the number of comments for each pattern.
A value of 0 expresses perfect equality (\ie each comment pattern consists of the same number of comments).
A value of 1 expresses maximal inequality among values (\ie a few patterns capture many comments, and the remaining comments are spread into many single-comment patterns).

Let us consider the example of a specific human commenter in our dataset. This human made 73 comments belonging to 12 patterns. 9 of these patterns have exactly one comment. The other ones correspond to ``\emph{LGTM}'' (37 comments), ``\emph{\#\#Fixes\{Number\}}'' (22 comments) and ``\emph{lgtm}'' (5 comments).
As a result, the Gini coefficient for this commenter is very low 0.04, since most patterns (9 out 12) have the same number of comments.
Let us compare this to a bot in our dataset with a similar number of comments (61) and comment patterns (10). The number of comments in each pattern is more unequally distributed, ranging from 1 to 49 comments per pattern, a consequence of much more repetitive messages. As a result, its Gini coefficient is much higher, namely 0.52.

\begin{figure}[h]
	\centering
	\includegraphics[width=\columnwidth]{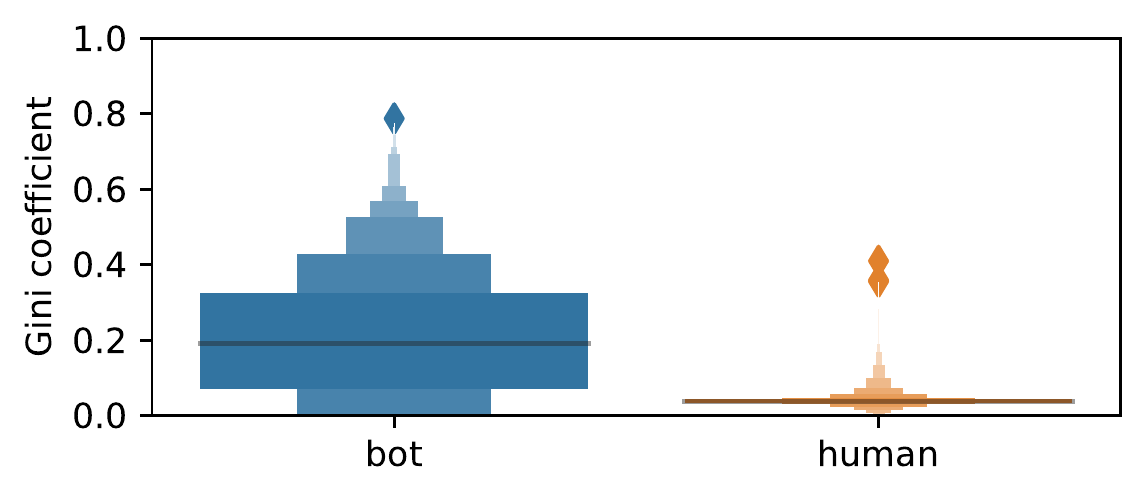}
	\caption{Distribution of Gini coefficient for bot and human commenters.}
    \label{ginidistribution}
\end{figure}

\fig{ginidistribution} shows the distribution of the Gini coefficient for all bots and humans in our dataset, by means of boxen plots~\cite{Lettervalueplots}.
We observe that humans exhibit a lower inequality than bots with respect to the spread of comments within patterns.
We statistically compared these distributions using a Mann-Whitney-U test~\cite{mann1947test}.
The null hypothesis, stating that the two distributions are the same was rejected ($p < 0.001$), indicating a statistically significant difference between the two distributions. The effect size turned out to be {\em large} (Cliff's delta $|d| = 0.58 $)~\cite{Cliff1993494,romano2006exploring}.
This confirms that humans tend to have a lower inequality than bots, a consequence of many of their patterns containing a single comment. Therefore, the Gini coefficient can help in distinguishing between bots and humans.

\subsection{Number of comments and empty comments}
\label{sec:comments}

In addition to the number of patterns and the unequal distribution of comments within patterns, we also consider the number of comments made by each commenter as a feature for our model.
This feature makes it possible to distinguish between commenters having a similar number of patterns. Indeed, consider for example two commenters having exactly 10 patterns. Assume they have respectively 10 and 100 comments. The first commenter is likely to be a human (since it has 10 patterns each containing exactly one comment, \ie all comments are different), while the second one is more likely to be a bot.

We also consider the number of empty comments as a feature for our model.
Indeed, during the rating process we found that a non-negligible proportion (6.5\%) of the considered comments were empty.
The presence of such comments in the dataset may seem strange.
Even if the \github user interface does not allow empty comments in a discussion, it does not prevent comments to be composed of white characters. Moreover, the \github user interface allows the creation of pull requests whose description is empty. Since this description is the very first comment of a pull request, it explains why we found empty comments in the dataset.

Interestingly, we found that empty comments are mostly created by human commenters and not by bots. For instance, only 7\% of all bots generated at least one such comment, whereas this proportion reaches 41.2\% for human commenters.
This should not come as a surprise, since one could expect bots mainly to generate informative comments and, by definition, empty comments are uninformative.
Consequently, we decided to consider the number of empty comments as a feature of our classification model.

\medskip

In summary, based on the analysis in this section we decided to use four distinct features for commenters to train the classification model:
\begin{enumerate*}[label=(\roman*)]
	\item the number of comment patterns;
	\item the inequality between comments in patterns;
	\item the total number of comments for the commenter; and
	\item the number of empty comments.
\end{enumerate*}\section{Classification model}
\label{sec:model}

All the scripts and data used to carry out the experiments in this section are available in a replication package on:\\
{\footnotesize \url{https://github.com/mehdigolzadeh/IdentifyBots_ReplicationPackage}}

\subsection{Classifier selection}
\label{sec:classifiers}

\begin{figure}
    \centering
    \includegraphics[width=\columnwidth]{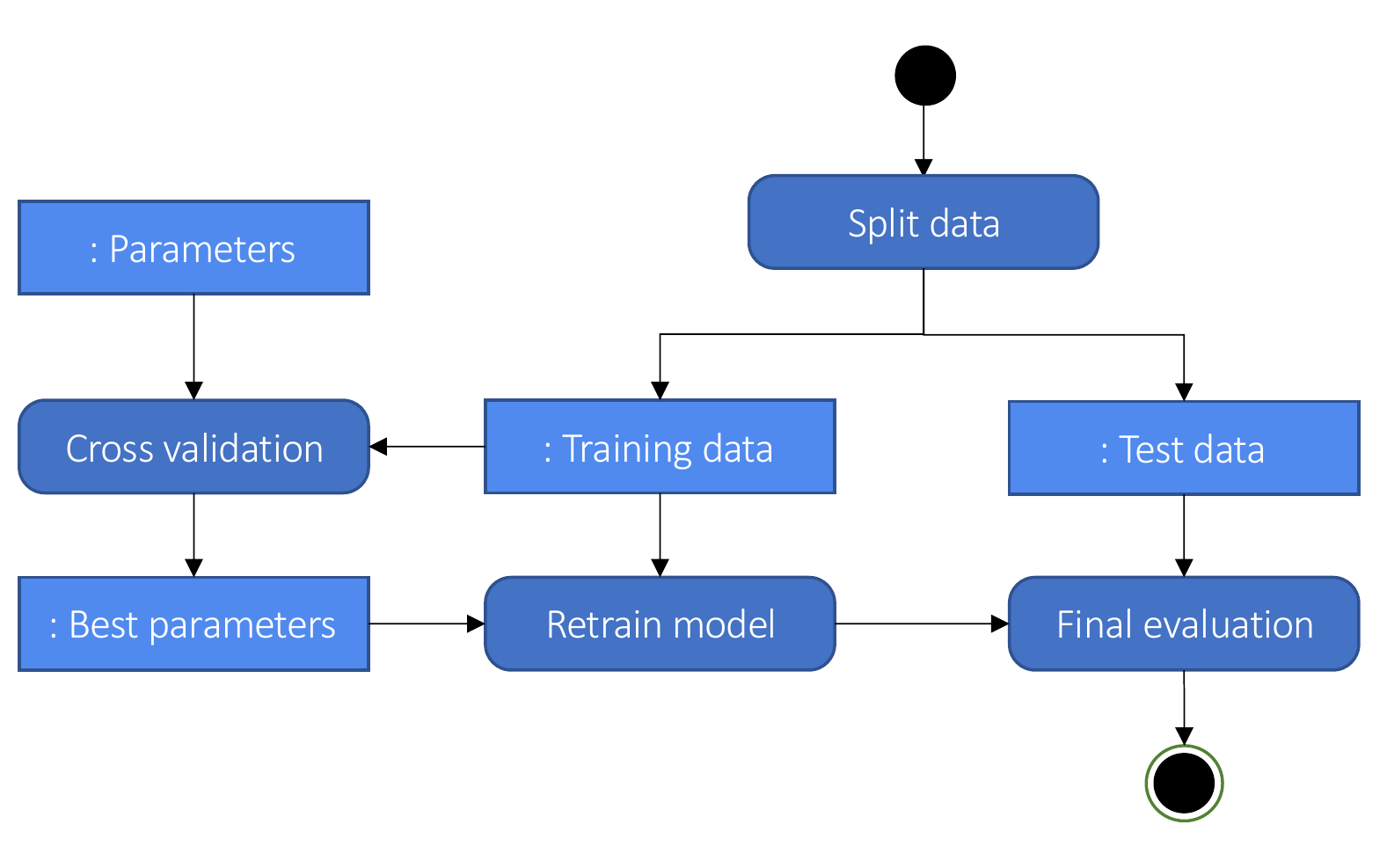}
    \caption{Standard workflow for grid-search cross-validation}
    \label{figworkflow}
\end{figure}

A wide variety of algorithms can be used to construct a classification model.
In this section we compare different classification algorithms to determine which one is the most appropriate to distinguish between bot and human commenters.
Among the classifiers having the ability to perform binary classification, we consider decision trees (DT) \cite{Safavian1991}, random forest (RF) \cite{breiman2001,Frank2001}, support vector machines (SVM) \cite{soton256459}, logistic regression (LR) \cite{McCullagh1991}, and k-nearest neighbours (kNN) \cite{Aha1991}.
Since the performance of these classifiers could depend on the input parameters, we follow a standard workflow of hyper-parameter tuning using a grid-search cross-validation process~\cite{Witten2011} (see \fig{figworkflow}).
To do so, we rely on \textsf{scikit-learn}~\cite{scikitlearn}, a well-known machine learning library for Python.

We first divided the ground-truth dataset into two disjoint sets: a training set containing 60\% of the data that will be used in a grid-search cross-validation process to determine the best input parameters and the best classifier, and a test set composed of the remaining 40\% that will be used to evaluate the performance of the selected classifier and parameters on new data.
Since we have many more humans than bots in our datasets, we relied on a stratified train-test split method to create these two sets with the same ratio of bots and humans.

Selecting an appropriate model with the best possible parameters requires hyper-parameter tuning. Based on the supported parameters of each classifier, we implemented a grid-search process based on a limited set of values for each parameter. For example, DT and RF were evaluated by setting the split criterion to \emph{Gini} and \emph{entropy}, among others.
Doing so resulted in 91 different classifiers.
To address the class imbalance problem~\cite{He2009} and avoid affecting the performance of the classifiers~\cite{Grbac2013}, we rely on a cost-sensitive learning approach~\cite{Elkan2001}. Practically, this means we set the \emph{class weight} parameter in \textsf{scikit-learn} to \emph{balanced} for each supported classifier.

We then trained and evaluated the performance of all classifiers using a 10-fold cross-validation process.
This approach splits the dataset into $10$ subsets of equal size, and for each fold a model is trained using $9$ subsets and is evaluated on the remaining one.
The overall performance of the model is averaged from the performance of these 10 models.
To ensure that the created subsets preserve the same proportion of bots and humans as in the complete training set, we relied on a \emph{stratified shuffle split} to create them.

The performance of the resulting models is measured using the classical metrics of precision $P$, recall $R$ and $F1$-score.
\changed{We use these metrics for the population of each class (\ie for bots $B$ and humans $H$). For the entire population we computed the \emph{weighted} version of these metrics to take into account the class imbalance.}
We aim to achieve an as high $F1$-score as possible. Since our goal is to identify bots, we also strive to keep bot recall $R(B)$ high enough, given that the population of bots is significantly smaller than the population of humans, and that it is much easier and faster to recover from humans misclassified as bots than the opposite.
All these metrics are summarized in \tab{tab:formulas}, and are defined in terms of the number of \emph{true positives \textbf{TP}} (the number of bots that are correctly classified as such by the model), \emph{true negatives \textbf{TN}} (the number of humans that are correctly classified as such by the model), \emph{false positives \textbf{FP}} (humans that are wrongly classified as bots), and \emph{false negatives \textbf{FN}} (bots that are wrongly classified as humans).

\begin{table*}[]
    \centering
    \caption{Definitions of precision, recall and $F1$-score.}
    \label{tab:formulas}
    \begin{tabular}{c|ccc}
    \toprule
    \bf population        & \bf precision $P$ & \bf recall $R$ & \bf $F1$-score \\
    \midrule
    bots $B$   & $\frac{ TP }{ TP + FP}$ & $\frac{ TP }{ TP + FN}$ & $\frac{ 2 \times P(B) \times R(B) }{ P(B) + R(B)}$ \\
    humans $H$ & $\frac{ TN }{ TN + FN}$ & $\frac{ TN }{ TN + FP}$ & $\frac{ 2 \times P(H) \times R(H) }{ P(H) + R(H)}$ \\
    \midrule
    $B\cup H$ &  $\frac{P(B)\times |B| + P(H) \times |H|}{|B| + |H|}$ &  $\frac{R(B) \times |B| + R(H) \times |H|}{|B| + |H|}$ & $\frac{ 2 \times P \times R}{P + R}$\\
    \bottomrule
\end{tabular}

\end{table*}

Following the grid-search cross-validation process described above, we trained and obtained 91 classifiers. For each of them, we computed the resulting \emph{bot}, \emph{human} and \emph{overall} precision, recall and $F1$-score.
\tab{tab:gridsearch} reports on these metrics, in descending $F1$-score order.
To ease readability, rather than reporting on all 91 classifiers, we selected for each classifier category (\eg DT, RF, ...) the instance whose parameters resulted in the highest $F1$-score.
\changed{We also compared the precision, recall and $F1$-score of these classifiers against a baseline classifier, ZeroR. ZeroR is a very simple classifier that has no predictive power: it ignores the features and always predicts the majority class (\ie ``human'' in our case). We observe that all classifiers exhibit a high overall performance (in terms of precision, recall and $F1$) and surpass ZeroR by a wide margin.
}

\begin{table*}[t]
    \centering
    \caption{Precision, recall and $F1$-score of the best classifiers per family of classifiers (in descending order of $F1$-score).}
    \label{tab:gridsearch}

    \begin{tabular}{r|cc|cc|ccc}
        \toprule
        & \multicolumn{2}{c|}{\bf bots} & \multicolumn{2}{c|}{\bf humans} & \multicolumn{3}{c}{\bf overall}\\
        \bf classifier & $P(B)$ & $R(B)$ & $P(H)$ & $R(H)$ & $P(B\cup H)$ & $R(B\cup H)$  & $F1(B\cup H)$ \\
        \midrule
        \bf RF &  \changed{0.932} &  \changed{0.916} &  \changed{0.990} &  \changed{0.992} &  \changed{0.984} &  \changed{0.984} &  \changed{0.984} \\
        \bf kNN &  \changed{0.943} &  \changed{0.853} &  \changed{0.983} &  \changed{0.994} &  \changed{0.978} &  \changed{0.979} &  \changed{0.978} \\
        \bf SVM &  \changed{0.876} &  \changed{0.925} &  \changed{0.991} &  \changed{0.984} &  \changed{0.979} &  \changed{0.978} &  \changed{0.978} \\
        \bf DT &  \changed{0.882} &  \changed{0.884} &  \changed{0.986} &  \changed{0.985} &  \changed{0.975} &  \changed{0.974} &  \changed{0.974} \\
        \bf LR &  \changed{0.839} &  \changed{0.931} &  \changed{0.992} &  \changed{0.978} &  \changed{0.975} &  \changed{0.973} &  \changed{0.974} \\
        \bf \changed{ZeroR} &  \changed{-} &  \changed{0.000} &  \changed{0.893} &  \changed{1.000} &  \changed{0.798} &  \changed{0.893} &  \changed{0.843} \\
        \bottomrule
    \end{tabular}

\end{table*}

\changed{The overall scores for $R$, $P$ and $F1$ of all classifiers are consistently higher than the ZeroR baseline, and range between 0.974 and 0.984. Even though the best SVM and LR classifiers have higher bot recall $R(B)$ than the best RF classifier ($0.925$ and $0.931$ compared to $0.916$, respectively), the overall $R$, $P$ and $F1$ scores are highest for the RF classifier.}
We therefore decided to use the best RF classifier, which was obtained with the \changed{\emph{entropy}} split criterion, 10 estimators (\ie trees) and a maximum depth of 10 for these trees.

\subsection{Evaluation}
\label{sec:ModelEvaluation}

In this subsection, we aim to evaluate the actual performance of that model on data that were not used to train the model, \ie on new data contained in the test set.
Following the workflow presented in \fig{figworkflow}, we start by constructing a new classification model instance based on the selected RF classifier, its parameters, and the training set containing 60\% of the ground-truth dataset.

We evaluate and report the accuracy of the model based on the test set, corresponding to the remaining 40\% of the ground-truth dataset.
This test set includes 2,000 commenters, of which 1,789 are humans and 211 are bots. The evaluation results are reported in \tab{tab:classificationreport}.

\begin{table}[h]
    \caption{Evaluation of the classification model using the test set.}
    \label{tab:classificationreport}
    \centering

    \begin{tabular}{l|ccrrr}
        {}    &  classified & classified & P  &  R & F1 \\
        & as bot & as human & & & \\
        \midrule
        Bot      &   \changed{TP: 192} &  \changed{\textbf{FN: 19}} &   \changed{0.94}  &  \changed{0.91}   &   \changed{0.92} \\
        Human      &  \changed{\textbf{FP: 13}} & \changed{TN: 1776} &    \changed{0.99}   &   \changed{0.99}  &   \changed{0.99} \\
        \midrule
        weighted avg      &   & &    \changed{0.98}  &  \changed{0.98}   &   \changed{0.98}\\
    \end{tabular}
\end{table}

We see that most bots and humans are correctly classified by the model. For instance, only \changed{19} out of 211 bots %
were misclassified as humans (\textbf{FN}), and only \changed{13} out of 1789 humans %
 were misclassified as bots (\textbf{FP}).
The overall $F1$-score is very high (\changed{0.98}), a consequence of the high precision (\changed{0.98}) and high recall (\changed{0.98}) of the model.
Thanks to the fact that we have taken into account class imbalance \changed{during the training phase}, \changed{these high scores can also be observed} individually for each class, even if the precision and recall for bots is slightly lower than for humans.
These results confirm what we already observed in previous section, that is, the model is effective in identifying bots and humans.

Scrutinising all \changed{19} misclassified bots ({\bf\em{FN}}) we found that \changed{ten} of them were already problematic during the \changed{first step of the} manual rating process, where they were rated as either ``Human'' or ``I don't know'' by one of the raters. Moreover,
the final decision to classify them as bots during the discussion session among raters was based on additional information that is not available in the comments themselves, explaining why the model is not able to classify them correctly.

The model also misclassified \changed{13} humans.
The fact that the model misclassified these humans as bots is not surprising given that, during \changed{the first step of the rating process}, %
\changed{10} out of \changed{13} cases were manually rated as \emph{difficult} or \emph{very difficult},
\changed{2} cases as ``I don't know'' by both raters and \changed{one case was even rated as a bot by one of the raters}.
\sect{sec:discussion} provides a detailed analysis of these misclassified commenters.

\medskip

Since the model relies on features computed on comments to distinguish bots from humans, it is worthwhile to consider and measure the impact of the number of considered comments on the performance of the model.
In particular, we aim to identify the minimal number of non-empty comments required to reliably classify bots and humans.
To this end, we evaluated our model and computed the $F1$-score for commenters in the test set, grouped by their number of non-empty comments.

\begin{figure}
    \centering
    \includegraphics[width=\columnwidth]{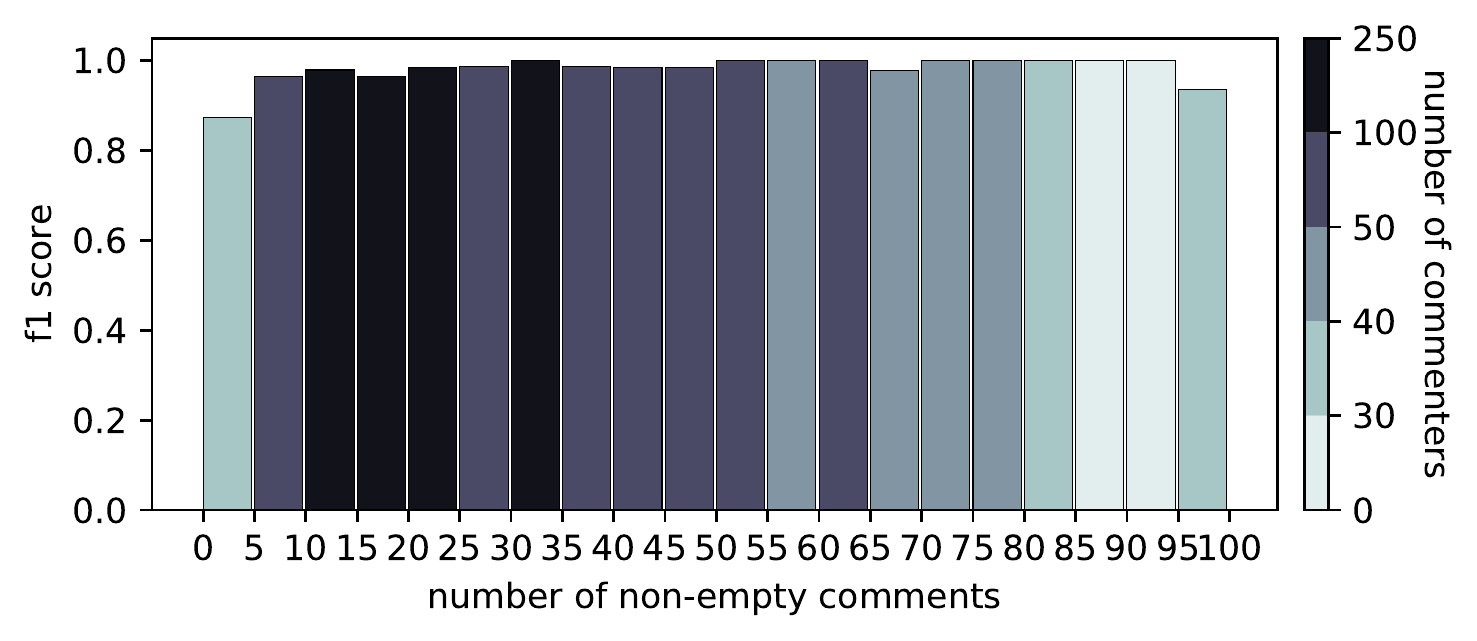}
    \caption{$F1$-score of the model when applied on commenters, grouped by their number of non-empty comments. The colour indicates the number of commenters in each bin.}
    \label{figsensitivity}
\end{figure}

\fig{figsensitivity} shows the resulting $F1$-scores of the model grouped by bins based on the number of non-empty comments.
The colour of a bin indicates how many commenters there are in that bin.
The bins with \changed{10 to 24 and 30 to 34} non-empty comments have the highest number of commenters, while bins between 85 to 94 have the lowest number of commenters.
The $F1$-score increases from \changed{0.87} (bin 0-4) and becomes stable around \changed{0.96 to 1.00 after 10} non-empty comments are reached (from bin 10-14).
This suggests that having at least 10 non-empty comments is enough to achieve good performance with the model.

\section{The \changed{\bodegha} bot detector tool}
\label{sec:tool}
Since the classifier we trained to identify bots presents very good performance, we implemented it as part of a tool. The tool is called \changed{\bodegha} (Bot Detector for GitHub activity), is developed for \python 3.7 and is easily installable through \textsf{pip}, the official package manager for \python.\footnote{Using \texttt{pip install git+https://github.com/mehdigolzadeh/BoDeGHa}}
\changed{\bodegha} can be used by any researcher or practitioner to classify accounts of a given \github repository either as bot or as human based on their issue and PR comments.

\begin{figure}
    \centering
    \includegraphics[width=\columnwidth]{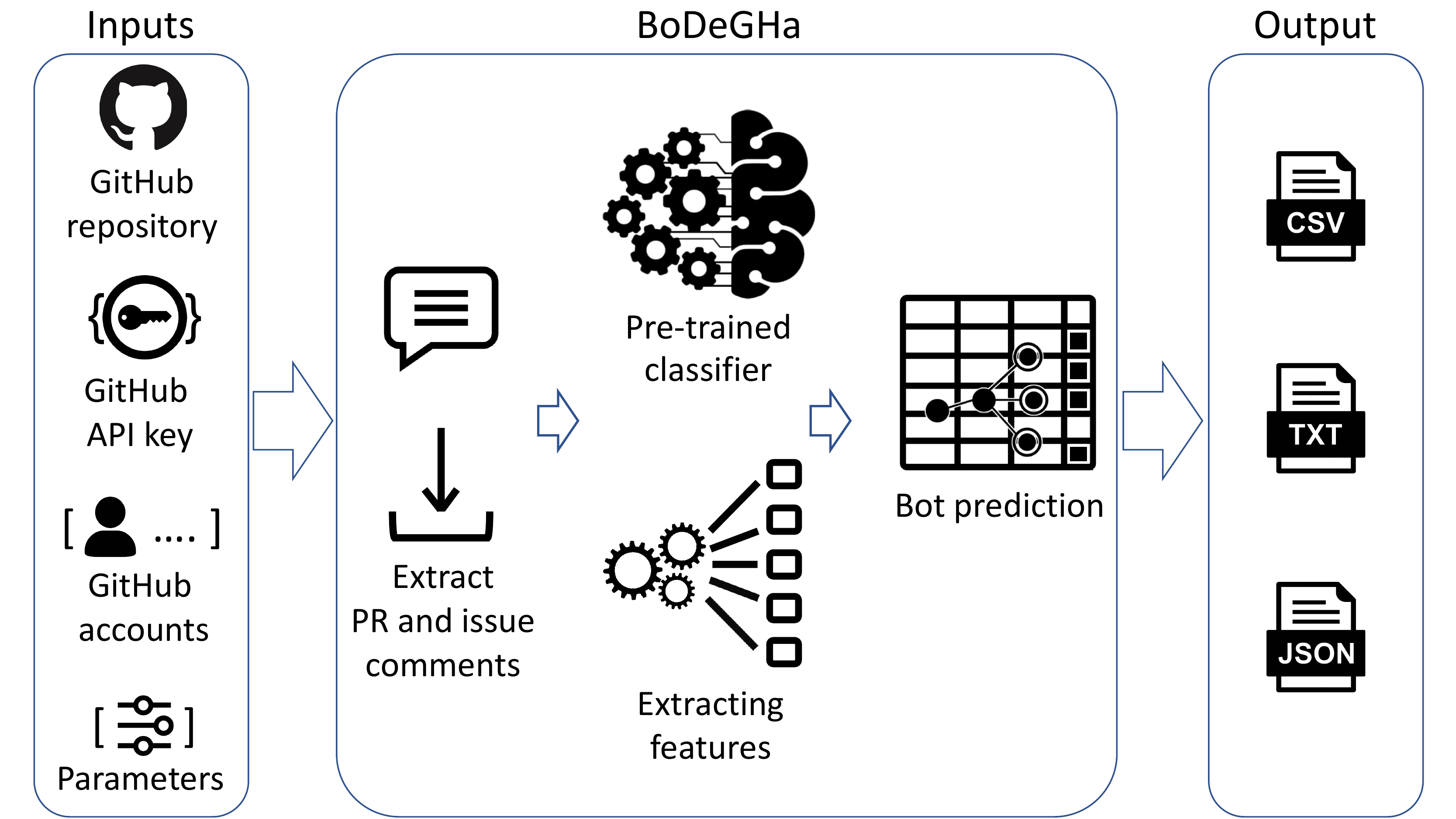}
    \caption{The \changed{\bodegha} architecture.}
     \label{fig:tool}
\end{figure}

In its simplest form, \changed{\bodegha} accepts the name of a \github repository and a \github API key.
\changed{\bodegha} computes its output in three steps, summarized in \fig{fig:tool}.
The first step consists of downloading all comments from the specified repository thanks to \github's GraphQL API. This step results in a list of commenters and their corresponding comments.
The second step consists of computing the number of comments, empty comments, comment pattern and inequality between number of comments within patterns (\ie the features of the classification model).
The third step simply applies the pre-trained model on these examples, and outputs the prediction made by the model.

\begin{figure}
    \centering
    \includegraphics[width=\columnwidth]{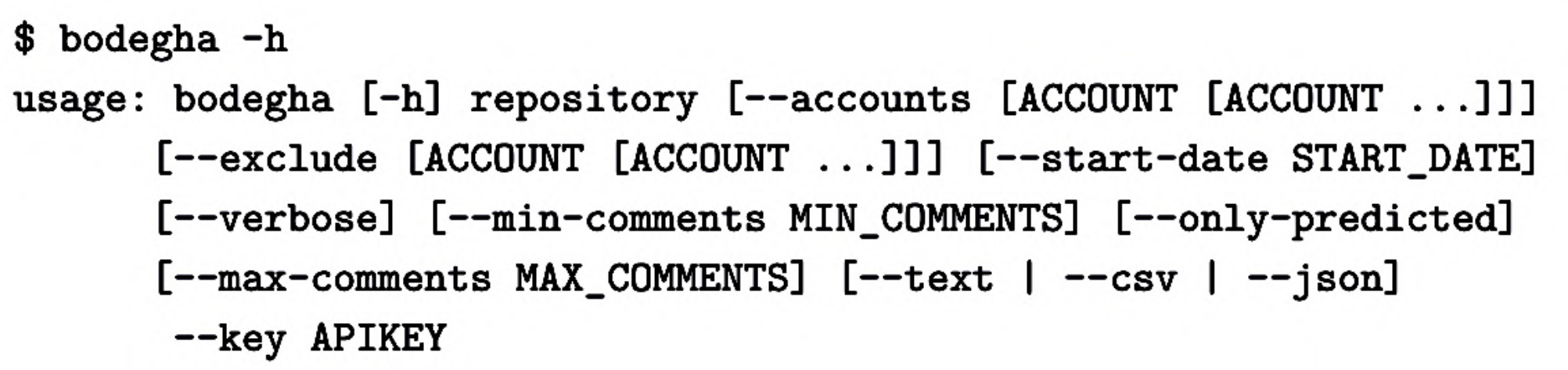}
    \caption{List of command-line arguments for \changed{\bodegha 1.0.1}.}
    \label{fig:command}
\end{figure}

\changed{\bodegha} supports several additional parameters. The minimum and maximum number of comments to download and to consider can be specified, as well as the start date from which to consider comments. It is also possible to provide a list of specific accounts for the tool to consider.
To ease its reuse by other tools, it is also possible to export the results either as comma-separated values or JSON.
The command-line interface of \changed{\bodegha} is summarized in \fig{fig:command}.

\begin{figure}
    \centering
    \includegraphics[width=\columnwidth]{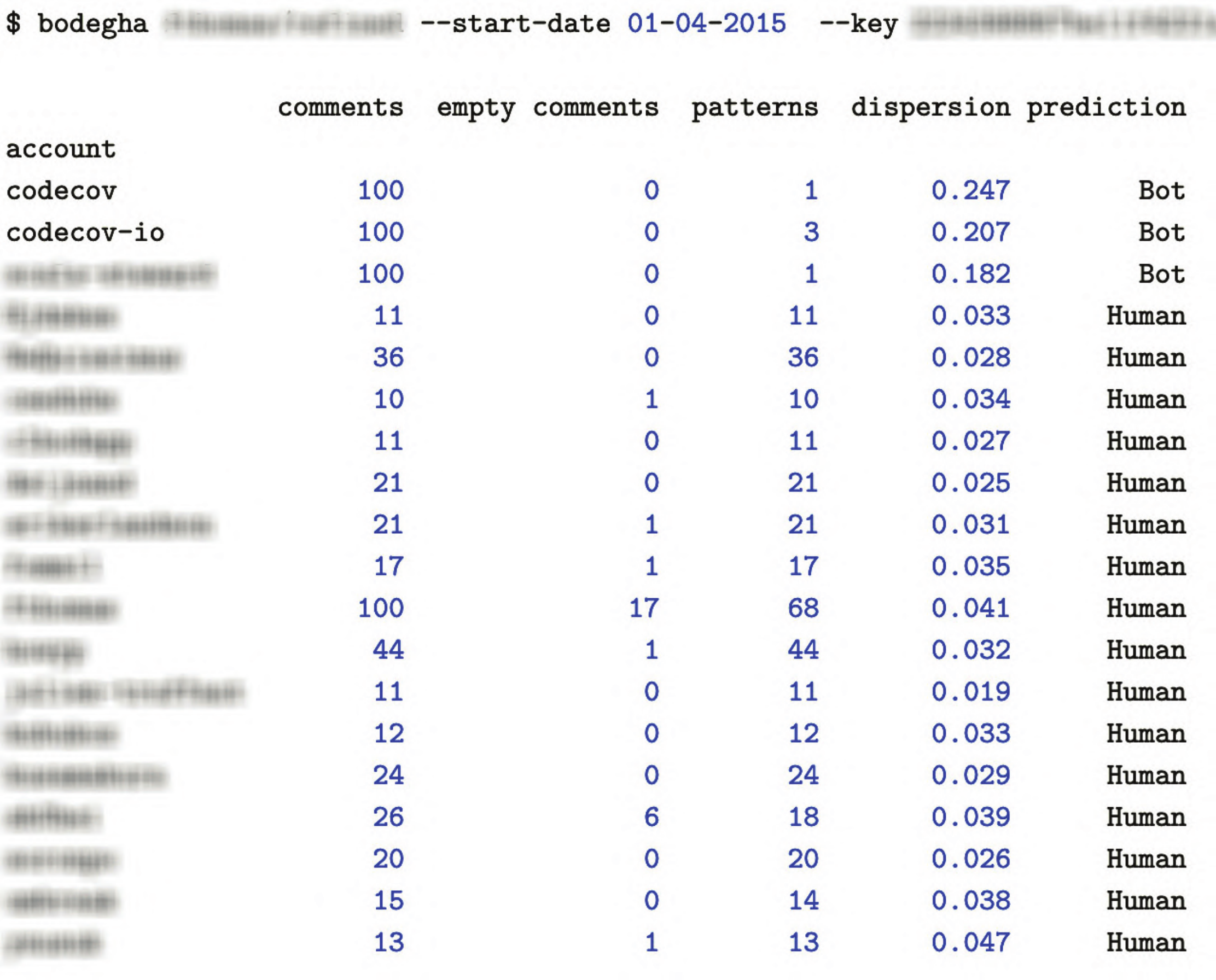}
    \caption{Example of running \changed{\bodegha}.}
    \label{fig:output}

\end{figure}

\fig{fig:output} presents the output of \changed{\bodegha} for a randomly chosen \github repository.
The output shows, for each \github account (first column), the number of extracted comments (second column), the number of empty comments (third column), the number of computed comment patterns (fourth column), and the inequality among them (fifth column). The last column provides the predicted class of each account.
This example shows that three commenters are identified as bots, and all remaining commenters as humans.

\section{Discussion}
\label{sec:discussion}

The evaluation of the classifier revealed several commenters that the model was not able to properly classify. We specifically look at the commenters that have been misclassified by the model. During the evaluation of the model on the test set, we found \changed{19} bots and \changed{13} humans that were misclassified. In order to have a more complete categorisation of misclassified commenters, we also applied the model on the training set and obtained \changed{4} additional bots and \changed{12} additional humans that are misclassified.

Starting with the \changed{24 (19+5)} bots, we found that in most cases they correspond to bots that use, convert or copy text that was initially produced by humans.
Even if these bots perform repetitive tasks (\ie copy information) and even if some of these bots use templates to transfer or copy comments that are recognizable to the human eye (\eg ``\emph{Jira issue originally created by user \{username\}: \{content of the issue\}}''), it is difficult for an automated algorithm to detect such cases.

\begin{description}[leftmargin=*]

\item[\emph{\changed{Copy from humans (9 bots):}}]
\changed{We found some instances of bots whose comments were generated based on content made by humans (\eg \textsf{taskcat-ci}, \textsf{trax-robot}). Since our model solely relies on features derived from comments, bot comments originating from human messages increase the likelihood of an incorrect classification.
Among these cases, we found several bots that transfer data (including issues, PRs and their associated comments) to \github from issue trackers, code review support tools, email etc.
For example,
\textsf{neos-bot} transfers all issues from a Jira issue tracker,
\textsf{suchabot} duplicates comments and issues from another system to \github,
and \textsf{wallabag-bot} migration from email content to \github.}

\item[\emph{\changed{Insufficient comments (9 bots):}}]
\changed{We found 9 bots (\eg \textsf{devtools-bot} and \textsf{egg-bot}) that were wrongly identified as humans due to the lack of a sufficient number of non-empty comments. Since our model relies on comment contents, bots with too few non-empty comments may lead to incorrect predictions even if these comments have similar comment patterns. We do not see any direct way to overcome this, since bots are expected to provide relevant information about what they are doing, and as such, one can expect their comments to be informative and non-empty.}

\item[\emph{\changed{Diverse comments (6 bots):}}]
\changed{We found 6 cases of bots that are used for the purpose of reporting, logging, or proposing code changes. The variation of comments in these bots increases the number of comment patterns, which prevents the model from identifying these bots. The source of the comment diversity
comes from the reports they send for each task. For example \textsf{sentry-io} creates an issue each time an error occurs in the software project, along with the details of this error (\eg stack trace). Another example is \textsf{violinist-bot} that submits a PR to update outdated dependencies and to report about the changes of this update. Despite these comments starting with a similar sentence (\eg ``\emph{Sentry Issue:}'' or ``\emph{If you have a high test coverage index, and your tests for this pull request are passing, it should be both safe and recommended to merge this update. Here is a list of changes between the version you use, and the version this pull request updates to:}''), they mainly consist of details related to the submitted issue or PR (\ie stack traces for \textsf{sentry-io} and list of issues for \textsf{violinist-bot}) and are considered as different comment patterns.
}
\end{description}

We also looked at the \changed{17 (13+4)} humans that were misclassified as bots, and created the following categories:\footnote{To comply with GDPR regulations, we cannot provide the account names for these cases.}

\begin{description}[leftmargin=*]

\item[\emph{Repetitive comments \changed{(8 humans)}:}]
We found \changed{8} instances of human commenters whose comments are mostly composed of repetitive messages, such as \emph{thank you} or \emph{LGTM} and that have nearly no other comments. Since repetitive messages are usually indicative of the presence of a bot, the model failed to correctly classify these commenters.

\item[\emph{\changed{Insufficient comments (3 humans)}:}]
We found \changed{3} humans with few comments, most of them being empty. Most of these comments were created in the context of a pull request whose title was already sufficiently informative.
Since these empty comments are grouped in a single comment pattern, and since they form the large majority of the comments made by these commenters, they were wrongly considered as being generated by a bot due to their repetitive nature. \changed{We also found instances where the comment content is too short or there are too few non-empty comments. This prompts our algorithm to group them into a small number of patterns and consequently provide wrong predictions.}

\item[\emph{Mostly unfilled issue templates \changed{(3 humans)}:}]
It is not unusual in \github repositories to require commenters  to follow a comment template or a checklist when creating issues or pull requests.\footnote{See \urlx{https://docs.github.com/en/github/building-a-strong-community/about-issue-and-pull-request-templates}} We found \changed{3} commenters whose comments were mostly composed of unfilled or barely filled templates, leading these comments to be considered as a single pattern, and leading the model to misclassify them as bots. Relying on an analysis of the content of such comments could prevent them from being misclassified, by taking into account the presence of such templates.

\item[\emph{Others \changed{(3 humans)}:}]
These cases do not fall into any of the above categories, and we have found no specific reason to explain their misclassification. Some of them have a small number of comments, while others only have a few patterns (\eg due to the presence of similar long URLs in comments) despite the fact that they do not seem to have duplicated or similar comments.
\end{description}

Most commenters that were misclassified by the classification model were also hard to recognize by the raters during the process of creating the ground-truth dataset.
In the test set, about \changed{84.6\% (11 out of 13)} of the humans that were misclassified as bots and about \changed{63.1\% (12 out of 19)} of the bots that were misclassified as humans were originally rated as ``I don't know'', ``difficult'', or ``very difficult'' by at least one of the raters. In contrast, among the correctly classified commenters, a much lower percentage of bots \changed{(12.5\%, 24 out of 192)} and humans \changed{(9.5\%, 169 out of 1772)} were rated as such.

Furthermore, during the creation of the ground-truth dataset, we encountered several examples of commenters whose features and comments were reminiscent of both humans and bots.
Such so-called ``mixed'' commenters are the result of \github accounts belonging to humans allowing automatic tools to use their account for carrying out certain specific tasks.
Hence, the comments of such commenters include both human-like and bot-like behaviour.
We identified 78 such commenters out of 5,082 commenters (\ie 1.5\%) during the rating phase and we consistently excluded them from the ground-truth dataset since we could not decide whether these commenters should be classified as bots or humans.

Nevertheless, it is interesting to report how our model behaves when exposed to these specific ``mixed'' cases.
Out of these 78 identified ``mixed'' commenters, \changed{21} were classified as bots \changed{(26.9\%)} and \changed{57 as a humans (73.1\%)}.
The fact that the proportion of ``mixed'' commenters classified as bots is higher than the one in the training set (\changed{10.3\%}) suggests that their behaviour is perceived to be closer to that of a bot than a human by the classification model.

\changed{The presence of mixed accounts as well as the categories of bots that have been misclassified as humans suggests that it is not easy to come up with a single definition for a bot. Two persons could easily disagree on whether a given account is a bot or a human if they have a different interpretation of what it means to be bot. This calls for a more precise definition of bots.
Erlenhov \etal~\cite{Erlenhov2020ESECFSE} started doing so based on qualitative interviews with developers.
This enabled them to identify three distinct DevBot {\em personas} that differ in terms of features like autonomy, chat interfaces, and smartness. This more fine-grained classification of DevBots and their characteristics paves the way for more sophisticated classification models.}

The approach presented in this paper is not the first one to have been proposed in the literature to detect bots in social coding platforms.
Dey~\etal~\cite{Dey2020MSR} proposed three different approaches for identifying bot accounts in \github projects, mostly based on their commit messages.
One of them consists of checking for the presence of the string ``bot'' in the account name of the committer.
We partially relied on this heuristic to add more potential bot candidates during our data collection.
However, solely relying on it to identify bots is likely to lead to a large number of both false positives and false negatives.
To confirm this, we applied their approach on our ground-truth dataset.
We found 169 humans out of 4,473 (3.8\%) containing the string ``bot'' in their account name, either at the end (46 cases) or in the middle (123 cases).
Out of the 527 bots we have in the dataset, 394 of them (\ie 74.7\%) actually contained ``bot'' in their account name, usually at the end of the name (378 cases).
Although this may seem high for such a simple heuristic, it still implies that more than \changed{one out of four bots} is missed with this method, and about \changed{one out of} 25 humans is mistakenly considered a bot.
For comparison, around only \changed{one out of} 25 (3.8\%) bots have been misclassified as humans by our model, and around only \changed{one out of} 100 humans (1.1\%).

\section{Threats to Validity}
\label{sec:threats}

Based on the structure recommended by Wohlin et al.~\cite{wholin2012} we discuss the threats that might call into question the validity of our findings, their potential impact and how we have tried to mitigate them.

\emph{Construct validity} examines the relationship between the theory behind the experiments performed and the observations found. This threat is mainly related to correctness of the dataset used in the experiments.
The results of our study are strongly dependent on the correctness of the ground-truth dataset. We are confident that the ground truth contains very few errors, since we
achieved an \emph{almost perfect} agreement ($\kappa=0.96$) based on an iterative rating process involving all authors of this paper.
One of the most likely threats is the existence of ``mixed'' commenters in the dataset.
Such commenters are difficult to classify, even by human raters, since they combine both bot-like and human-like behaviour.
Mixed commenters constitutes a very small proportion of our dataset (78 cases, corresponding to 1.5\% of all considered accounts).
We excluded all these cases from the dataset since we could not agree on them.
However, it is possible that the dataset still contains such cases that were not identified by the raters. %
Given the very low ratio of such mixed accounts, it is however unlikely to affect our findings.

\emph{Internal validity} concerns choices and parameters of the experimental setup that could affect the results of the observations.
Given that our classification method is fully based on features computed from comments, we required each commenter included in the dataset to have contributed at least 10 (possible empty) comments.
This threshold is based on previous experiences and findings~\cite{Golzadeh2020}.
As such, we cannot claim that our model applies on commenters who made fewer than 10 comments.
Similarly, we considered at most 100 comments for each commenter but, as explained in \sect{sec:ModelEvaluation}, this upper limit on the number of comments is unlikely to have biased our results, since we already achieved high $F1$-score starting from 10 non-empty comments.

\emph{Conclusion validity} concerns whether the conclusions derived from the analysis are reasonable.
Our conclusions are based on the evaluation and application of the classification model on the test set.
Given that we properly followed a standard grid-search cross-validation method to identify the best classifier, and that we evaluated the model on the test set (\ie examples that have not been used to train or select the classifier), the results we obtained and conclusions we reached are unlikely to be affected.

\emph{External validity} concerns the degree to which the conclusions we derived are generalisable outside the scope of this study.
The main threat to external validity is related to the construction of the ground-truth dataset. To avoid any potential bias, we randomly selected a large collection of \github repositories related to software development and corresponding to actual packages being officially distributed, following the guidelines of Kalliamvakou \etal~\cite{Kalliamvakou2014}.
While this dataset can be regarded as representative of bots contributing to \github repositories through PR and issue comments, we do not make any claim about its generalisability to other activities (\eg commit messages) or other social coding platforms (\eg BitBucket or GitLab).
Nevertheless, the underlying approach could be made applicable to such activities or platforms.

\section{Future work}
\label{sec:futurework}

As future work, we intend to use our classification model in socio-technical empirical analyses of collaborative software development, by studying the effect of the presence of bots on various development-related activities, such as the productivity and quality of handling issues, bugs and pull requests, code reviewing, intra- and inter-repository collaboration, developer onboarding, and so on.

With the emergence of more advanced AI, machine learning and natural language processing techniques, we can expect future bots to behave more and more like humans.
These new technological advances may make our classification model %
less capable of distinguishing bots from humans.
\changed{An example of such technique is \emph{Generative Pre-trained Transformer 3} (GPT-3), an autoregressive language model developed by \textsf{OpenAI} and integrating 175 billion parameters. GPT-3 generates text of sufficiently high quality that it is difficult to distinguish them from that written by humans~\cite{brown2020language}.}
To cope with this, we will explore more advanced machine learning methods that could take into account the semantics of the comments. In particular, we will consider techniques relying on natural language processing and deep neural networks to develop classification models that are more resilient to human-like bots, as well as ``mixed accounts'' corresponding to bots that copy, transfer or translate human comments. %

\changed{In this study, we provided a binary definition of commenters being either bots or humans. 
The aforementioned study by Erlenhov \etal~\cite{Erlenhov2020ESECFSE} revealed that a more fine-grained definition would be needed, and they came up with three DevBot {\em personas} based on their autonomy, chat interface, and smartness. 
Such a more fine-grained classification could be used to refine our ground-truth dataset, and to provide more advanced classification models, possibly even computing the probability that an account belongs to each of the considered {\em personas}.}

Because of the growing use of bots during collaborative development activities~\cite{Erlenhov2019}, we can expect to see a proliferation of bots to automate software development in \github repositories.
For instance, \github introduced in November 2019 \github Actions\footnote{\url{https://github.com/features/actions}}, a feature providing automated workflows for repository maintainers.
These actions, fully integrated with \github, allow the automation of tasks based on a various set of triggers (\eg commits, pull request, issue, comments, etc.).
Since they are easily shareable from one repository to another one (through the \github Marketplace), we expect their use to become more widespread, even in smaller repositories and, as a result, to see more ``bots'' and their comments in \github repositories.
However, tasks triggered through \github actions are automatically labelled as such by the \github API, eliminating the need to create a model to identify these ``bots''.
Recently, \github action variants of many well-known bots (\eg \textsf{Coveralls}, %
\textsf{Codecov}, %
\textsf{Snyk}) %
have been published to the \github Marketplace, and these actions are rapidly increasing in popularity.
Consequently, we expect their \github action variant to replace progressively the bots currently being used in \github repositories.

\changed{We aim to extend our study, including the ground truth dataset, classification model and tool to accommodate other social coding platforms such as \gitlab and \bitbucket. This will generalise our approach, and allow us to study to which extent the selected platform affects the way in which bots are being used as part of the development process. We also aim to evaluate our model on other types of activities (\eg git commit messages). This will allow us to understand to what extent our model can be used to identify bots based on commit activities.}
\section{Conclusion}
\label{sec:conclusion}

In this paper, we proposed a novel approach to distinguish between bots and humans in collaborative software development repositories on \github, based on the comments they made in issues and PRs.

Our first contribution is the creation of a ground-truth containing 5,000 \github accounts including 527 bots (10.5\%), based on a manual rating process with very high inter-rater agreement ($\kappa =0.96$).

Using this ground-truth dataset, we developed a classification model to identify bots based on four features: the total number of comments of a commenter; its number of non-empty comments; its number of comment patterns; and the inequality between the numbers of comments in each pattern.
The chosen features align with behavioural differences we observed between bots and humans. Indeed, we found that most human commenters tend to have diverse sets of comments with little repetition, while bots tend to frequently use a limited set of comment patterns.

Following a standard grid-search 10-fold cross validation process, we evaluated and compared five families of classifiers (random forest, k-nearest neighbours, decision trees, logistic regression and support vector machines) on a training set including 60\% of all data. We performed hyper-parameter tuning to select the best parameters of each classifier family based on their precision, recall and $F1$-score. We selected the random forest classifier since it achieved the highest $F1$-score (\changed{98.4\%}).

We evaluated the selected classifier on new data, and found that it achieves \changed{high} precision and recall. %
Based on a manual assessment and categorisation of bots and humans that were misclassified, we identified why the classification model had difficulties with detecting them, and we provided suggestions for further improvements to the classification model.

We implemented the classification model into a \python command-line tool, called \changed{\bodegha}. This open source tool is made freely available to practitioners and researchers to allow them to analyse \github repositories and to identify which accounts correspond to bots and which correspond to humans.

\bibliographystyle{cas-model2-names}
\bibliography{botsbiblio}

\end{document}